\documentclass[doublecol]{epl2} 

\usepackage{graphicx}
\usepackage{hhline}
\usepackage{dcolumn}
\usepackage{bm}

\title{Profile and scaling of the fractal exponent of percolations in complex networks}
\shorttitle{Profile and scaling of the fractal exponent of percolation in network} 

\author{T. Hasegawa\inst{1} 
\thanks{E-mail:  \email{hasegawa@m.tohoku.ac.jp}} 
\and T. Nogawa\inst{2} 
\and K. Nemoto\inst{3} \thanks{E-mail:  \email{nemoto@statphys.sci.hokudai.ac.jp}} 
}
\shortauthor{T. Hasegawa \etal}

\institute{                    
  \inst{1} 
Graduate School of Information Sciences, Tohoku University, 6-3-09, Aramaki-Aza-Aoba, Sendai, 980-8579, JAPAN \\
  \inst{2} 
Faculty of Medicine, Toho University, 5-21-16, Omori-nishi, Ota-ku, Tokyo 143-8540, JAPAN \\
  \inst{3} 
Department of Physics, Graduate School of Science, Hokkaido University, Kita 10-jo Nisi 8-tyome, Sapporo, JAPAN
}
\pacs{64.60.ah}{First pacs description}
\pacs{64.60.aq}{Second pacs description}
\pacs{89.75.Hc}{Third pacs description}

\abstract{
We propose a novel finite size scaling analysis for percolation transition observed 
in complex networks.
While it is known that cooperative systems in growing networks often undergo 
an infinite order transition with inverted Berezinskii-Kosterlitz-Thouless singularity, 
it is very hard for numerical simulations to determine the transition point precisely. 
Since the neighbor of the ordered phase is not a simple disordered phase but a critical phase, 
conventional finite size scaling technique does not work. 
In our finite size scaling, the forms of the scaling functions for the order parameter and the fractal exponent 
determine the transition point and critical exponents numerically for an infinite order transition 
as well as a standard second order transition. 
We confirm the validity of our scaling hypothesis through Monte-Carlo simulations for bond percolations in some network models: 
the decorated (2,2)-flower and the random attachment growing network, where an infinite order transition occurs, 
and the configuration model, where a second order transition occurs.
}

\begin{document}

\maketitle

\section{Introduction}

The science of complex networks has led to new perspectives in the study of statistical physics~\cite{albert2002statistical,newman2003structure,boccaletti2006complex,barrat2008dynamical}.
Various processes, such as percolations, epidemics, spin systems, 
and coupled oscillators, in complex networks have been actively studied (see~\cite{dorogovtsev2008critical} and references therein).

An elementary theoretical framework of critical phenomena 
in complex networks has been provided with the local tree approximation~\cite{dorogovtsev2008critical}.
Equilibrium systems on uncorrelated networks with arbitrary degree distribution $P(k)$ ($k$ denotes degree), 
built with the configuration model \cite{molloy1995critical}, 
are well described by using the local tree approximation. 
In particular, the critical properties for the uncorrelated network 
with scale-free degree distribution $P(k) \propto k^{-\gamma}$ 
are determined only by the exponent $\gamma$~\cite{newman2003structure,dorogovtsev2008critical}.

However, the systems on networks made with growth mechanism show a quite different picture.
Let us consider the bond percolation on networks, 
where each bond is open with probability $p$. 
The bond percolations on Euclidean lattices and uncorrelated networks 
show a conventional second order phase transition 
between the non-percolating phase, where only finite size clusters exist, 
and the percolating phase, where there exist a unique infinite cluster and finite size clusters.
On the other hand, analytical studies for several stochastic and deterministic growing networks have revealed that 
the system exhibits an unusual phase transition, termed infinite order transition with inverted Berezinskii-Kosterlitz-Thouless (BKT) singularity
~\cite{callaway2001randomly,dorogovtsev2001anomalous,lancaster2002cluster,kim2002infinite,coulomb2003asymmetric,zalanyi2003properties,pietsch2006derivation,rozenfeld2007percolation,zhang2008degree,boettcher2009patchy,berker2009critical,hasegawa2010critical,hasegawa2010generating} 
(see also~\cite{bauer2005phase,khajeh2007berezinskii,hinczewski2006inverted,boettcher2011fixed,nogawa2012generalized,nogawa2012criticality}
for the case of spin systems):
(i) The singularity of the phase transition is infinitely weak. 
When $p$ is larger than the transition probability $p_c$, the order parameter $m(p) = \lim_{N \to \infty} S_{\rm max}(p, N)/N$, 
where $S_{\rm max}(p, N)$ is the size of the largest cluster averaged over percolation trials in the system with $N$ nodes, follows 
\begin{equation}
m(p) \propto \exp [-\alpha/(\Delta p)^{\beta'}] \,, \label{infiniteorder}
\end{equation}
where $\Delta p=p-p_{c}$.
(ii) Below the transition point, the mean number of clusters with size $s$ per node $n_s$ obeys the power law,
\begin{equation}
n_s \propto s^{-\tau} \,. \label{ns}
\end{equation}
%

%
\begin{table*}[!t]
\caption{Expected scenario of phase transitions of bond percolations on various graphs.
The number of ends of an infinite graph $G$ is defined as 
the supremum of the number of infinite connected components in 
$G \backslash S$, where $G \backslash S$ is the graph obtained from $G$ 
by removing arbitrary finite subgraph $S$ and the edges incident to those~\cite{lyons2000phase,schonmann2001multiplicity}.} 
\label{relation}
\begin{tabular}{llc}
\hline graph type & example & scenario \\
\hline \hline amenable graph with two ends 
& chain 
& $0<p_{c1}=p_{c2}=1$ \\ \hline 
amenable graph with one end & $d(\ge 2)$-dimensional Euclidean lattice & $0<p_{c1}=p_{c2}<1$\\ \hline 
NAG with infinitely many ends & Cayley tree & $0<p_{c1}< p_{c2}=1$ \\ \hline 
NAG with one end & enhanced binary tree \cite{NAG,NAG2,NAGcomment,baek2012upper,gu2012crossing}, 
hyperbolic lattice~\cite{baek2009hyperbolic,lee2012bounds,gu2012crossing} & $0<p_{c1}<p_{c2}<1$ \\ \hline 
stochastic growing tree & growing random tree~\cite{hasegawa2010critical,zhang2008degree,lancaster2002cluster,zalanyi2003properties,pietsch2006derivation} & $0=p_{c1}<p_{c2}=1$ \\ \hline 
stochastic growing network model & RAGN ($m$-out graph)~\cite{zalanyi2003properties,riordan2005small,bollobas2005slow}, 
CHKNS model~\cite{dorogovtsev2001anomalous,callaway2001randomly} & $0=p_{c1}<p_{c2}<1$ \\ \hline 
deterministic growing network & decorated (2,2)-flower~\cite{rozenfeld2007percolation,berker2009critical,hasegawa2010generating}, 
HN5~\cite{boettcher2009patchy} & $0=p_{c1}<p_{c2}<1$ \\
& 
HN-NP~\cite{boettcher2009patchy,hasegawa2013absence}, Boettcher-Singh-Ziff network~\cite{boettcher2012ordinary} & 
\\
\hline \end{tabular} 
\end{table*}
%
%

As indicated in ~\cite{hasegawa2010critical,hasegawa2010generating}, 
the above unusual phase may be same as 
the critical phase (the intermediate phase \cite{lyons2000phase,schonmann2001multiplicity}) observed in nonamenable graphs (NAGs).
NAGs are defined as infinite graphs with positive Cheeger constant~\cite{lyons2000phase,schonmann2001multiplicity}.
The Cheeger constant $h(G)$ of an infinite graph $G$ is given as $h(G)={\rm inf}_K |\partial K|/|K|$, 
where $K$ is an arbitrary nonempty subset of $V(G)$, which is the set of nodes in $G$, 
and $\partial K$ consists of all nodes in $V(G)-K$ that have a neighbor in $K$.
Typical examples of NAGs are hyperbolic lattices and trees.
The bond percolation on a NAG takes three distinct phases~\cite{benjamini1996percolation,lyons2000phase,schonmann2001multiplicity}:
(i) the non-percolating phase ($0 \le p < p_{c1}$),
(ii) the critical phase ($p_{c1} \le p \le p_{c2}$) in which there are infinitely many infinite clusters, 
and (iii) the percolating phase ($p_{c2} < p \le 1$).

The scenarios of percolations 
on some types of graphs are summarized in table~\ref{relation}.
The phase boundaries of amenable graphs including the Euclidean lattices, 
and of NAGs satisfy 
$0 < p_{c1}=p_{c2} \le 1$ and $0 <p_{c1}<p_{c2} \le 1$, 
respectively~\cite{lyons2000phase,schonmann2001multiplicity}.
If the above-mentioned unusual phase on growing networks 
is regarded as the critical phase on NAGs, 
we may say that the growing networked systems have a new scenario:
$p_{c1}=0$ and $p_{c2} > 0$ (last three rows in table~\ref{relation}).
However, we have no methodology to predict numerically which relation of the three (or none of them) 
holds for unknown networks.

For finite size systems, 
we usually focused on the estimation value of the largest or mean cluster size.
In recent studies~\cite{NAG,NAG2,hasegawa2010critical,hasegawa2010generating,boettcher2012ordinary,hasegawa2013absence}, 
the fractal exponent $\psi$ has been introduced to distinguish the above three phases.
The fractal exponent is defined as 
\begin{equation}
\psi(p) = \lim_{N \to \infty} \psi(p, N), \quad 
\psi(p, N) = \frac{{\rm d} \ln S_{\rm max}(p, N)}{{\rm d} \ln N}, 
\end{equation} 
which mimics $d_f/d$ for $d$-dimensional Euclidean lattice systems, 
$d_f$ being the fractal dimension of the largest clusters. 
In the non-percolating phase, $\psi(p, N)$ goes to zero 
when $N$ increases because $S_{\rm max}(p, N)$ remains finite.
In the percolating phase, $\psi(p, N)$ approaches one with $N \to \infty$ because $S_{\rm max}(p, N) \propto N$.
In the critical phase, $\psi(p, N)$ converges to a certain value between 0 and 1 when $N$ is large, and varies with $p$~\cite{NAG}. 
Because $n_s \propto s^{-\tau}$ in this phase, 
$\psi$ plays a role of a natural cutoff exponent of $n_s$. 
Then we have $\tau=1+\psi^{-1}$ from $N \int_{N^\psi}^\infty {\rm d} s n_s \sim O(1)$
~\cite{hasegawa2010critical, hasegawa2010generating}.

Finite size scaling analysis is a powerful method to extract the transition point and 
critical exponents in the large size limit from numerical data.
It is indeed the case for 
some static networks such as the configuration model, 
as reported in~\cite{hong2007finite}.
For growing networks, however, we are faced with the following two difficulties 
in the estimation of the transition point $p_{c2}$ from critical behaviors. 
First, we should perform a scaling analysis with data only for $p \ge p_{c2}$, 
and but not for $p < p_{c2}$. 
This is because the transition at $p_{c2}$ takes place 
between the percolating phase and the critical phase instead of the non-percolating phase. 
The system is always critical for $p<p_{c2}$, 
and there does not exist any characteristic scale that grows to diverge at $p_{c2}$. 
Therefore, the standard scaling analysis cannot be applied~\cite{hasegawa2010critical}.
Secondly, the transition at $p_{c2}$ is often of infinite order, 
for which the standard scaling theory based on power-law does not work.

In this letter, we propose a novel finite size scaling analysis 
to investigate numerically both the growing and non-growing networked systems.
Our scaling form is described 
in terms of the network size $N$ and the fractal exponent $\psi$.
As an application, we evaluate the fractal exponent and perform the scaling analysis 
for the some exactly-solved cases: 
the decorated (2,2)-flower~\cite{rozenfeld2007percolation,berker2009critical,hasegawa2010generating} 
and the random attachment growing network~\cite{zalanyi2003properties}
(RAGN; a random attachment version of the Barab\'asi-Albert model~\cite{barabasi1999emergence}), 
both of which exhibit an infinite order transition with inverted BKT singularities~\cite{zalanyi2003properties,riordan2005small,bollobas2005slow},
and on the configuration model having the same degree distribution with the RAGN, 
which exhibits a second order transition between the non-percolating and the percolating phases.
Our finite size scaling analysis works well for all of these cases.

\begin{figure*}[ttttt]
\begin{center}
\includegraphics[trim=25 5 60 20,scale=0.22,clip]{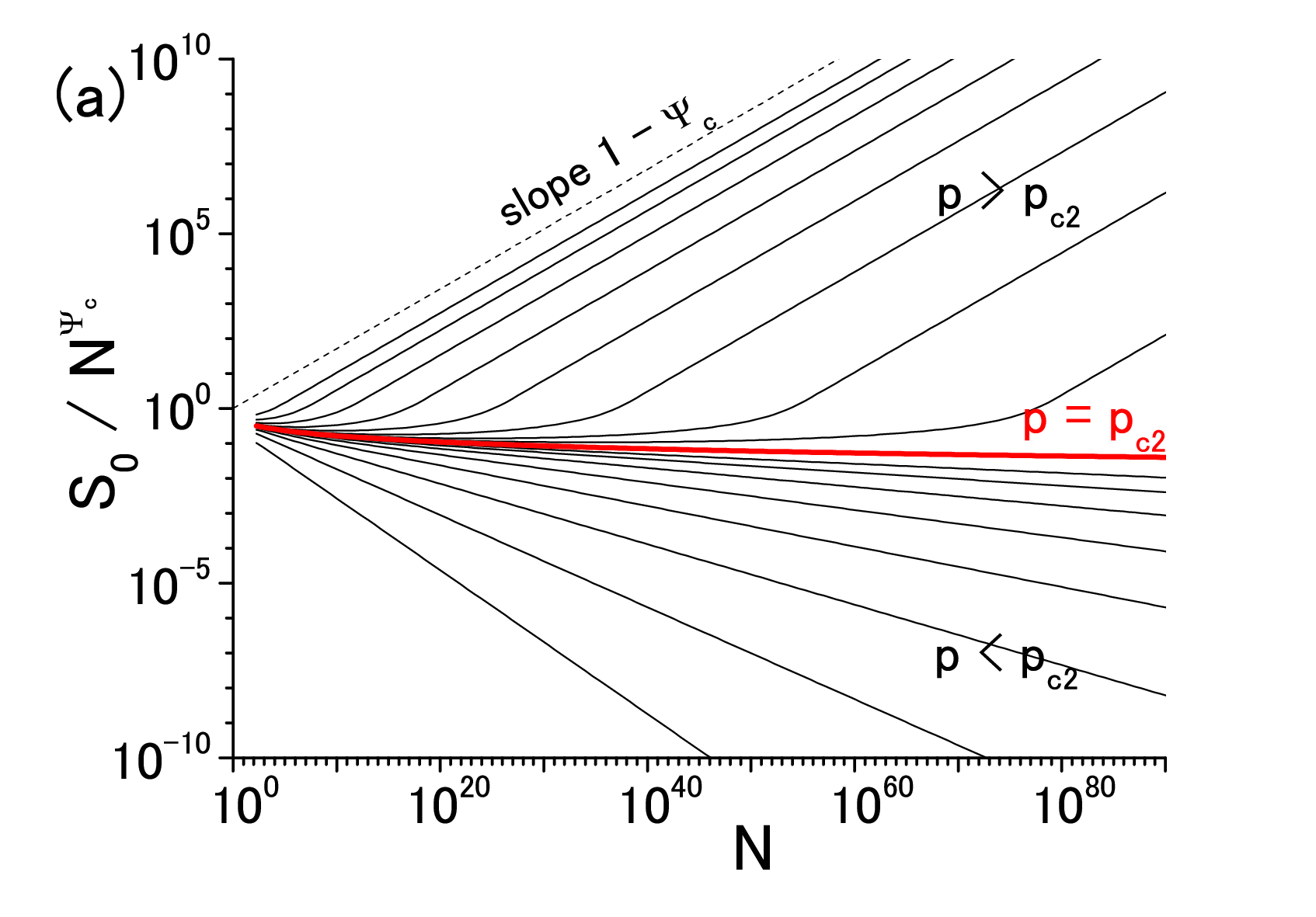}
\includegraphics[trim=25 5 60 20,scale=0.22,clip]{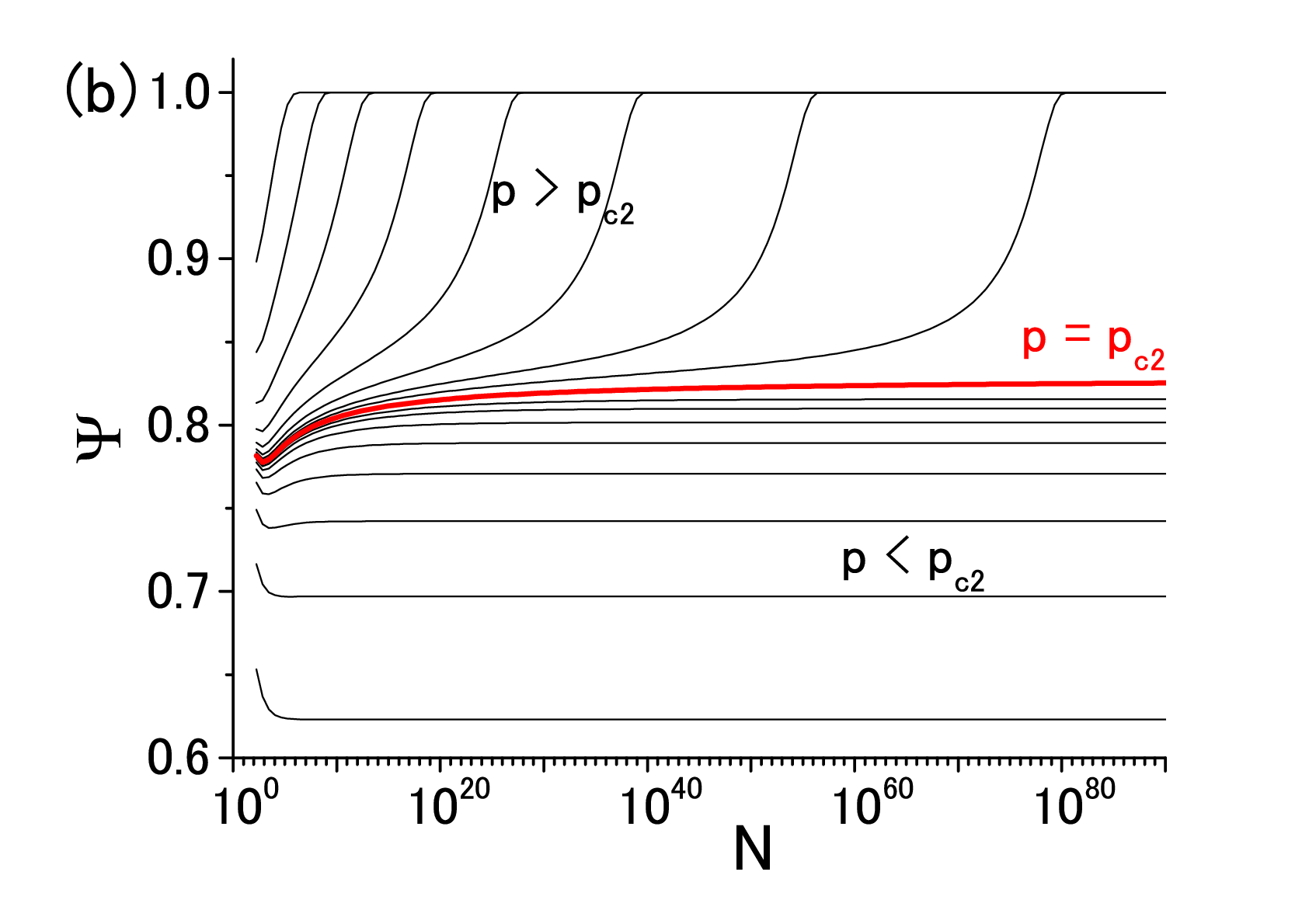}
\includegraphics[trim=25 5 60 20,scale=0.22,clip]{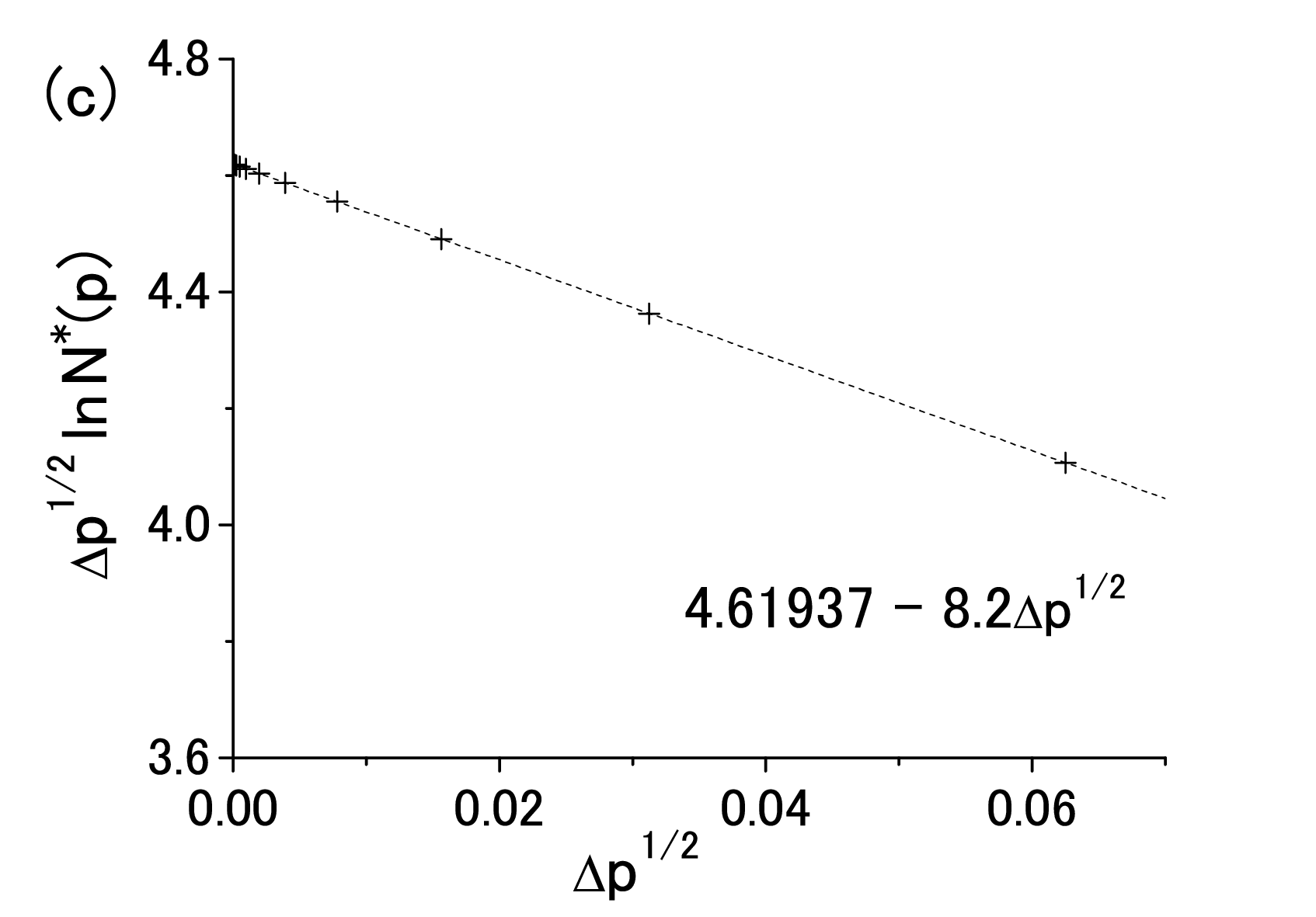}
\end{center}
\caption{ 
Size dependence of (a) $S_0$ and (b) $\psi$ of the decorated (2,2)-flower. 
(c) The dependence of the crossover scale $N^*$ at which $\psi(p, N)=0.95$ on $\Delta p = p - p_{c2}$ for $p > p_{c2}$. 
}
\label{flowerdata1}
\end{figure*}

\section{Scaling argument}

Let us heuristically derive a scaling form with $\psi$ and $N$ for $p>p_{c2}$.
First, we consider the $N$ dependence of $S_{\rm max}(p, N)$ at $p$ slightly larger than $p_{c2}$.
We suppose that there exists a crossover size $N^*(p)$, which diverges as approaching $p_{c2}$. 
For $N \ll N^*(p)$, the system behaves as if it were critical even for $p > p_{c2}$, 
so that $S_{\rm max}(p, N) \propto N^{\psi_c}$, $\psi_c$ being the fractal exponent at $p=p_{c2}$.
For $N \gg N^*(p)$, we obtain the behavior in thermodynamic limit as $S_{\rm max}(p, N)=Nm(p)$.
By connecting these two limits at $N=N^*$, we expect $N^*(p)$ as
\begin{equation}
N^*(p) \propto m(p)^{1/(\psi_c-1)} \,, \label{crossover}
\end{equation}
and the finite size scaling form for $S_{\rm max}(p, N)$ as 
\begin{equation}
S_{\rm max}(p, N) =N^{\psi_c} f_1 \Big[ \frac{N}{N^*(p)} \Big] \,, 
\label{SmaxScalingForm}
\end{equation}
where scaling function $f_1$ satisfies 
\begin{equation}
f_1(x) \propto 
{\Biggl\{}
\begin{array}{ccl}
{\rm const} & {\rm for} & x \ll 1 \\
x^{1-\psi_c} & {\rm for} & x \gg 1 \,. 
\end{array} \label{SmaxScaling1}
\end{equation}
Or equivalently, 
\begin{equation}
S_{\rm max}(p, N) =N^*(p)^{\psi_c} f_2 \Big[ \frac{N}{N^*(p)} \Big] \,, 
\label{SmaxScalingForm2}
\end{equation}
where
\begin{equation}
f_2(x) = x^{\psi_c} f_1(x) \propto 
{\Biggl\{}
\begin{array}{ccl}
x^{\psi_c} & {\rm for} & x \ll 1 \\
x & {\rm for} & x \gg 1 \,.
\end{array} \label{SmaxScaling2}
\end{equation}
By taking the logarithmic derivative of eq.~(\ref{SmaxScalingForm2}) with $N$, 
we obtain a finite size scaling form for $\psi(p, N)$ as 
\begin{eqnarray}
\psi(p, N) 
= g \Big[ \frac{N}{N^*(p)} \Big] \,, 
\label{psiScalingForm}
\end{eqnarray}
where
\begin{equation}
g(x) =\frac{{\rm d} \ln f_2(x)}{{\rm d} \ln x}=
{\Biggl\{}
\begin{array}{ccl}
\psi_c & {\rm for} & x \ll 1 \\
1 & {\rm for} & x \gg 1 \,.
\end{array} \label{psiScaling}
\end{equation}
Similarly, the finite size scaling form for the average size of clusters, 
$S_{\rm av}(p, N)=\sum_{s \neq S_{\rm max}} s^2 n_s$, can be assumed as
\begin{equation}
S_{\rm av} (p, N) = N^{\psi_{\rm av}} h \Big[ \frac{N}{N^*(p)} \Big] \,, 
\label{chiScalingForm}
\end{equation}
where 
\begin{equation}
h(x) =
{\Biggl\{}
\begin{array}{ccl}
{\rm const} & {\rm for} & x \ll 1 \\
x^{-\psi_{\rm av}} & {\rm for} & x \gg 1 \,,
\end{array} \label{chiScaling}
\end{equation}
$\psi_{\rm av}$ is the fractal exponent of the mean cluster size at $p=p_{c2}$, defined as $S_{\rm av}(p_{c2}, N) \propto N^{\psi_{\rm av}}$ 
and is related to $\psi_c$ as $\psi_{\rm av} = 2 \psi_c-1$ 
because $S_{\rm av}(p_{c2}, N) \sim \int^{S_{\rm max}(p_{c2}, N)} {\rm d}s s^2 n_s \propto N^{2\psi_c-1}$.

In the case of a second order transition, $m(p) \propto (\Delta p)^\beta$ for $p \ge p_{c2}$($=p_{c1}$).
Then the finite size scaling forms for $S_{\rm max}(p, N)$, $\psi (p, N)$, and $S_{\rm av}(p, N)$ are 
\begin{equation}
S_{\rm max}(p, N) = N^{\psi_c} f_1 [ N (\Delta p)^{\beta/(1-\psi_c)}] \,, 
\label{SmaxScalingForm-recon}
\end{equation}
\begin{equation}
\psi(p, N) = g [ N (\Delta p)^{\beta/(1-\psi_c)}] \, , 
\label{PsiScalingForm-recon}
\end{equation}
and
\begin{equation}
S_{\rm av}(p, N) = N^{\psi_{\rm av}} h [ N (\Delta p)^{\beta/(1-\psi_c)}] \,, 
\label{SavScalingForm-recon}
\end{equation}
respectively.

In the case of an infinite order transition, $m(p)$ follows eq.~(\ref{infiniteorder}). 
Then, we obtain the finite size scaling form of infinite order transition for $S_{\rm max}(p, N)$ and $\psi(p, N)$ 
by substituting eq.~(\ref{crossover}) into eqs.~(\ref{SmaxScalingForm}) and (\ref{psiScalingForm}) as 
\begin{equation}
S_{\rm max}(p, N) =N^{\psi_c} f_1 ( N \exp [-\alpha/(1-\psi_c)(\Delta p)^{\beta^\prime}]) \,, 
\label{SmaxScalingFormInfiniteOrder}
\end{equation}
and  
\begin{equation}
\psi(p, N)= g ( N \exp [-\alpha/(1-\psi_c)(\Delta p)^{\beta^\prime}]) \,, 
\label{PsiScalingFormInfiniteOrder}
\end{equation}
respectively.

As seen above, the character of singularity is encapsulated in $N^*(p)$, 
and various kind of singularity would be treated in the same framework by changing the function form of $m(p)$.
Similar relation for equilibrium spin systems was derived by renormalization group analysis for a hierarchical small-world network \cite{nogawa2012generalized}.

Note that the present scaling form includes the conventional finite size scaling. 
For $d$-dimensional lattice systems, the fractal exponent satisfies $\psi_c=1-\beta /d \nu$, 
so that eq.~(\ref{SmaxScalingForm}) reduces to the conventional scaling for $N m$ 
provided that $N/N^*(p) =(L/\xi)^d$, 
where $L$ is the linear dimension, and $\nu$ the critical exponent of correlation \textit{length} $\xi \propto (\Delta p)^{-\nu}$.

\section{Numerical check}

To check the validity of our scaling argument, 
we perform the numerical calculations for some exactly solved models: the decorated (2,2)-flower, the RAGN and the configuration model.

\begin{figure}[t]
\begin{center}
\includegraphics[trim=20 0 20 20,scale=0.24,clip]{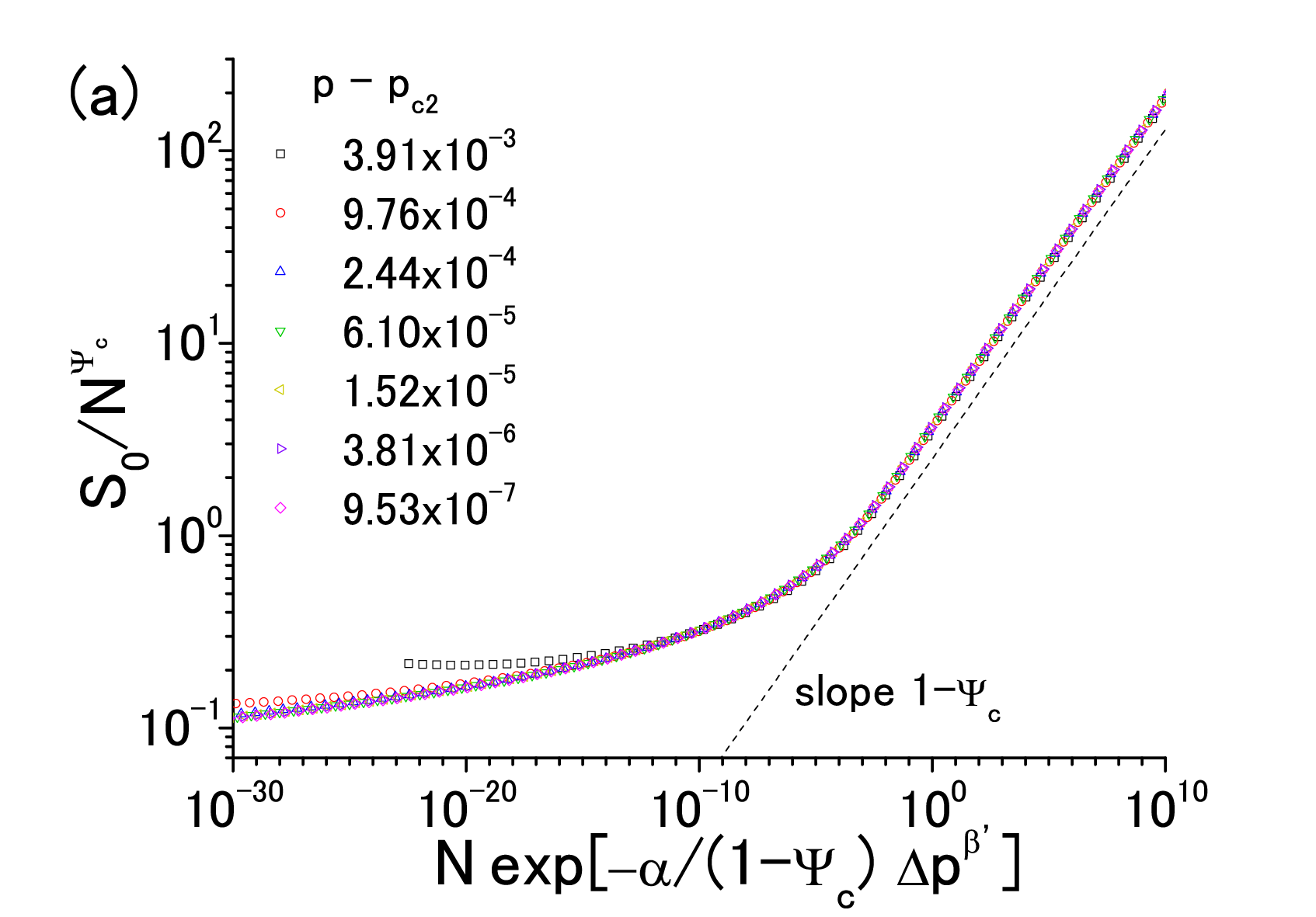}
\includegraphics[trim=20 0 20 20,scale=0.24,clip]{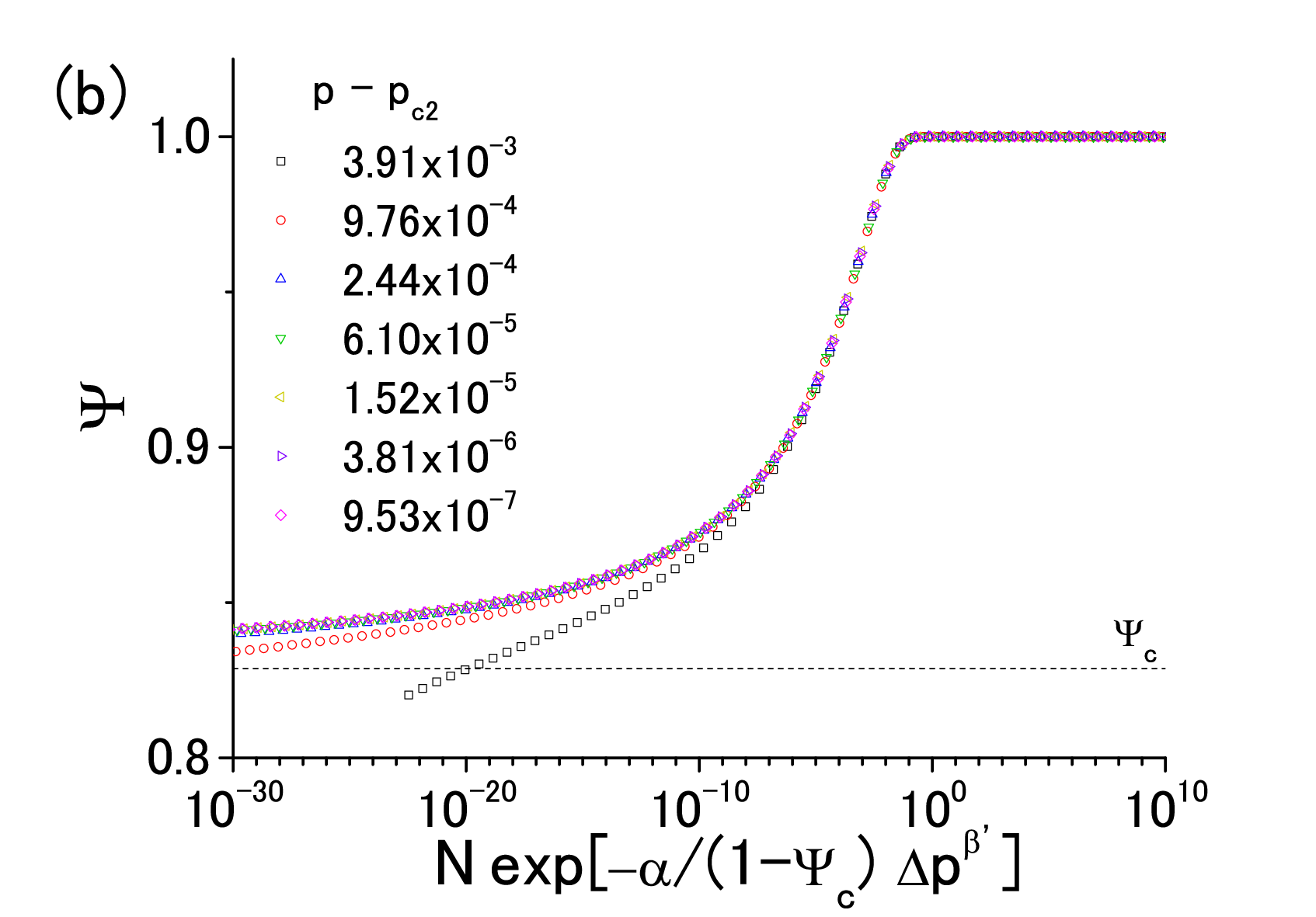}
\end{center}
\caption{ 
Scaling plot of (a) $S_0(p, N)$ and (b) $\psi(p, N)$ 
of the decorated (2,2)-flower, for $p>p_c$. 
Here we used $p_{c2}=5/32$, $\psi_c = 1/2 + \log_4 ( 1 + \sqrt{3} )$, $\beta'$=1/2, and $\alpha$ = 0.791049. 
}
\label{flowerdata2}
\end{figure}


\begin{figure}[t]
\begin{center}
\includegraphics[width=6.5cm]{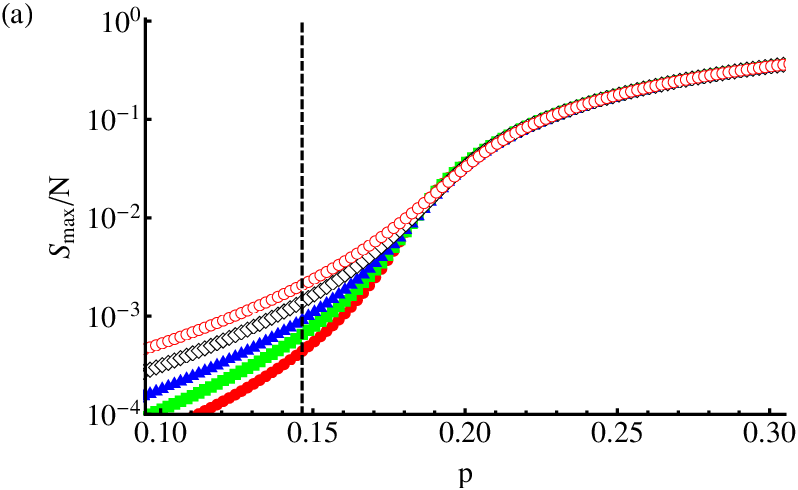}
\includegraphics[width=6.5cm]{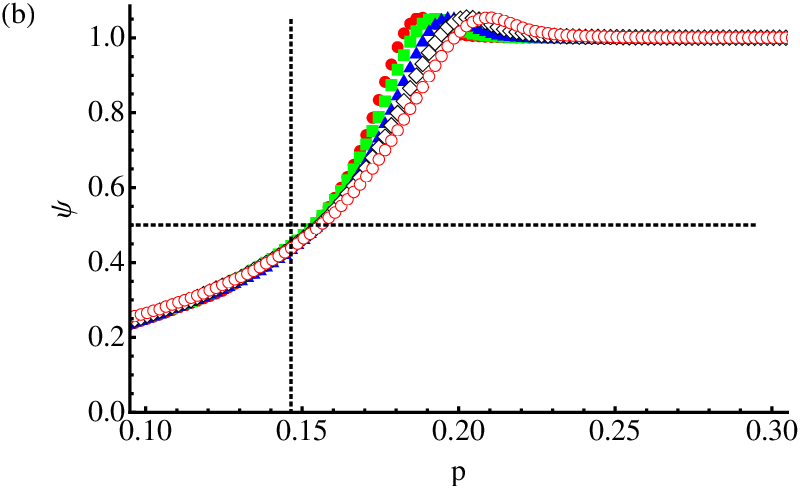}
\end{center}
\caption{ 
(a) The order parameter $S_{\rm max}(p, N)/N$ and (b) the fractal exponent $\psi(p, N)$ of the RAGN. 
The numbers of nodes $N$ are 
$2^{20}$ (red-circle), $2^{19}$ (green-square), $2^{18}$ (blue-triangle), $2^{17}$ (open-diamond), and $2^{16}$ (open-circle).
The vertical dashed line indicates $p_{c2}=(1-1/\sqrt{2})/2$, 
and the horizontal dashed line indicates $\psi_c = 1/2$.
}
\label{rawdata}
\end{figure}

\begin{figure}[!h]
\begin{center}
\includegraphics[width=6.5cm]{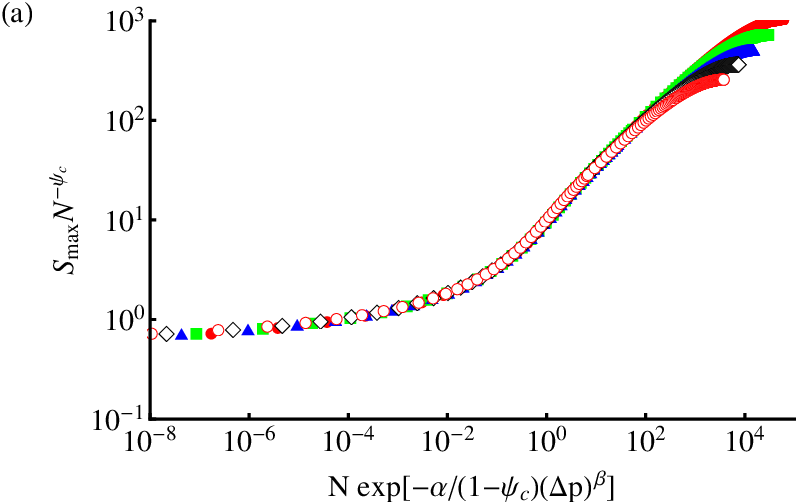}
\includegraphics[width=6.5cm]{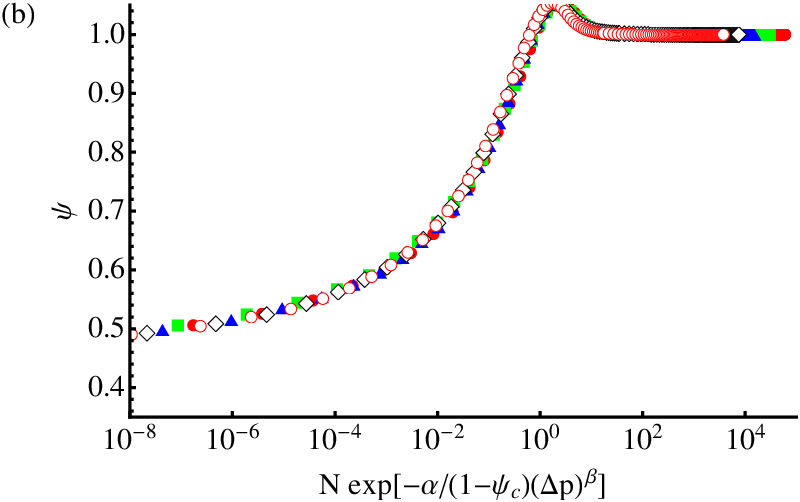}
\end{center}
\caption{
Finite size scaling for 
(a) $S_{\rm max} (p, N)$ by eq.~(\ref{SmaxScalingFormInfiniteOrder}) 
and (b) $\psi (p, N)$ by eq.~(\ref{PsiScalingFormInfiniteOrder}) of the RAGN.
$N=2^{20}$ (red-circle), $2^{19}$ (green-square), $2^{18}$ (blue-triangle), $2^{17}$ (open-diamond), and $2^{16}$ (open-circle).
}
\label{scaling}
\end{figure}


\subsection{decorated (2,2)-flower}

The decorated (2,2)-flower $F_n$ with generation $n$ is constructed in a deterministic way~\cite{rozenfeld2007percolation}.
Starting from $F_0$ which consists of two nodes (referred to as the {\it roots}) connected by a bond, 
$F_n$ is obtained from $F_{n-1}$, such that each edge added newly in $F_{n-1}$ creates a node which is connected to both ends of its edge. 
The number of nodes in $F_n$ is $N=N_n=2(4^n+2)/3$.

Bond percolation on the decorated flower can be solved by renormalization group technique~\cite{rozenfeld2007percolation,berker2009critical,hasegawa2010generating}.
The transition points are $p_{c1}=0$, $p_{c2}=5/32$ and $\psi_c = 1/2 + \log_4 ( 1 + \sqrt{3} ) \approx 0.82875$.
We can calculate the mean size of the cluster including one of the roots (the root cluster) $S_0(p, N)$ 
by the recursive equations of the generating functions (eqs.~(7-9,11) in \cite{hasegawa2010generating}).
Assuming that the fractal exponent $\psi_0(p)$, such that $S_0 (p, N) \propto {N}^{\psi_0(p)}$, is equivalent to $\psi(p)$ \cite{hasegawa2010generating}, 
we consider $S_0(p, N)$ instead of $S_{\rm max}(p, N)$ to check the validity of our scaling argument.

We plot $S_0 (p, N) /N^{\psi_c}$ and $\psi(p, N)$ with respect to $N$ at several values of $p$ in figs.~\ref{flowerdata1}(a) and (b), respectively.
Here we define $\psi(p, N_{n}) = \log_4 [S_0(p, N_{n})/S_0(p, N_{n-1})]$.
For $p<p_{c2}$, $S_0 (p, N)$ is proportional to ${N}^{\psi(p)}$ with $\psi(p)<\psi_c$ for large $N$. 
For $p>p_{c2}$, $S_0 (p, N)$ shows crossover around certain $N^*(p)$ as mentioned in eqs.~(\ref{SmaxScalingForm})-(\ref{SmaxScaling2}).
Similarly, $\psi(p, N)$ converges to $\psi(p)$ for $p<p_{c2}$, and shows stepwise change from $\psi_c$ to 1 for $p>p_{c2}$.
Here we calculate $N^*(p)$ as a value of $N$ at which $\psi(p, N) = 0.95$. 
Its $p$-dependence is shown in fig.~\ref{flowerdata1}(c), which is consistent with our hypothesis 
$N^*(p) \propto e^{\alpha/(1-\psi_c) \Delta p^{\beta'} }$, i.e., $\Delta p^{\beta'} \ln N^*(p) = \alpha/(1-\psi_c) + {\rm const.} \times \Delta p^{\beta'}$ by supposing $\beta' = 1/2$. 
From this plot we obtain $\alpha = 0.791049$.


\begin{figure*}[!t]
\begin{center}
\includegraphics[width=5.8cm]{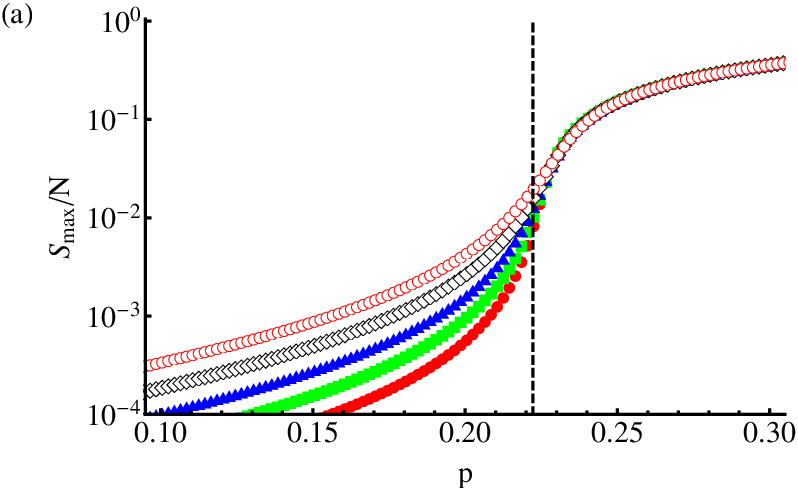}
\includegraphics[width=5.8cm]{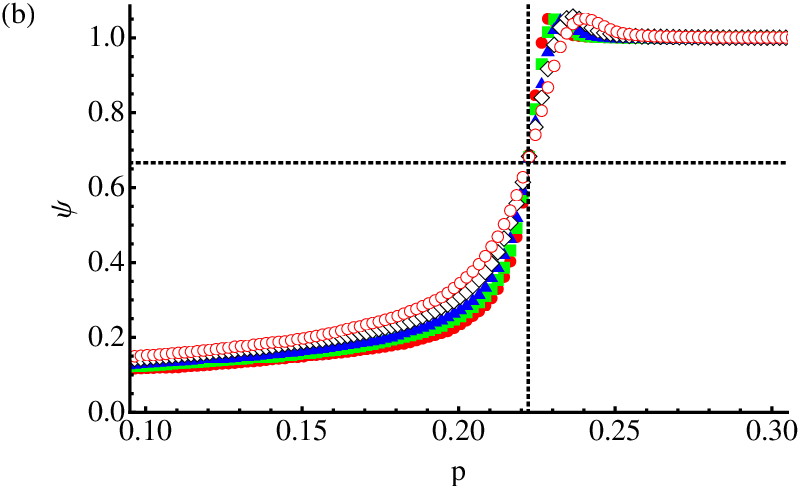}
\includegraphics[width=5.8cm]{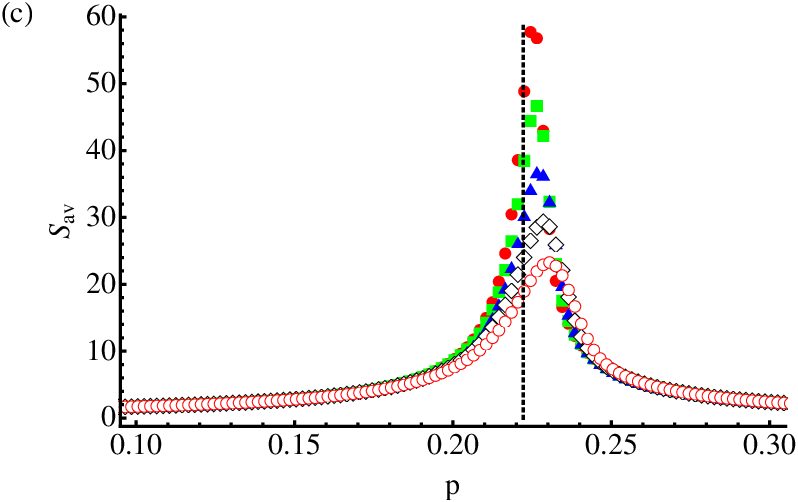}
\end{center}
\caption{
(a) The order parameter $S_{\rm max}(p, N)/N$, (b) the fractal exponent $\psi(p, N)$, 
and 
(c) the mean cluster size $S_{\rm av}(p, N)$ of the configuration model.
The numbers of nodes $N$ are 
$2^{20}$ (red-circle), $2^{19}$ (green-square), $2^{18}$ (blue-triangle), $2^{17}$ (open-diamond), and $2^{16}$ (open-circle).
The vertical dashed line indicates $p_{c2}=2/9$, 
and the horizontal dashed line indicates $\psi_c=2/3$.
}
\label{reconrawdata}
\end{figure*}

\begin{figure*}[!t]
\begin{center}
\includegraphics[width=5.8cm]{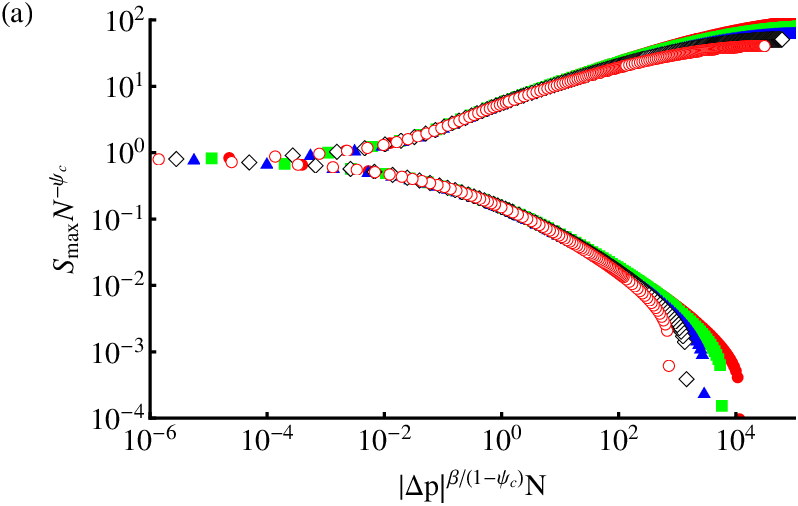}
\includegraphics[width=5.8cm]{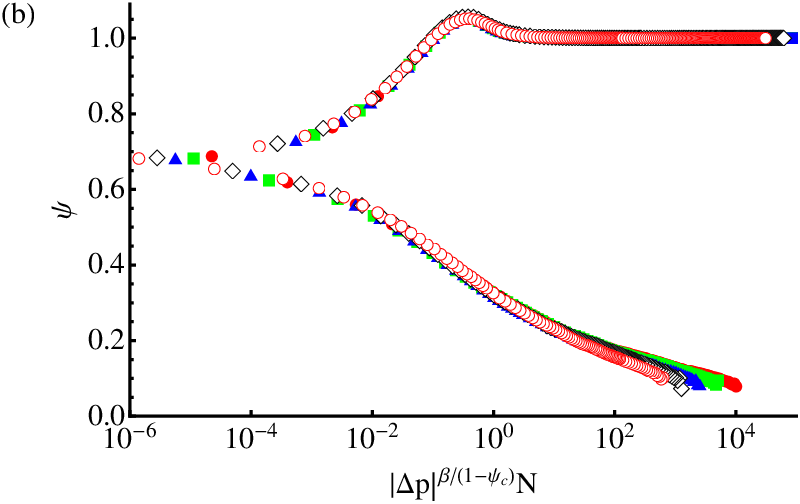}
\includegraphics[width=5.8cm]{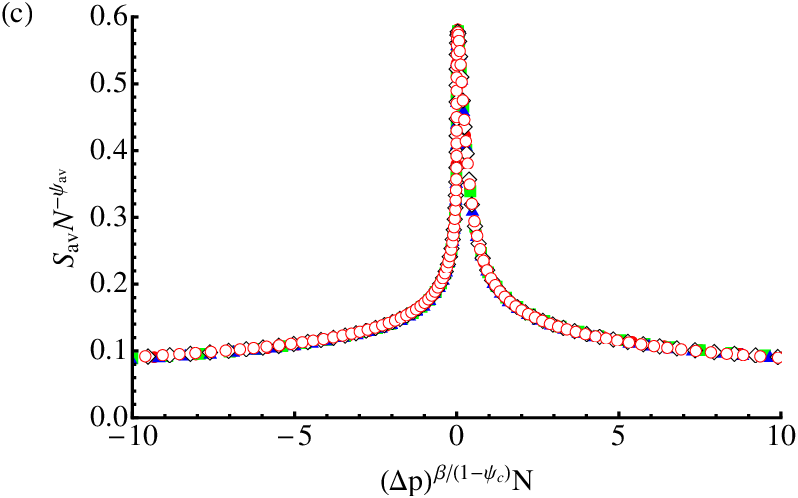}
\end{center}
\caption{
Finite size scaling for 
(a) $S_{\rm max}(p, N)$ by eq.~(\ref{SmaxScalingForm-recon}), 
(b) $\psi (p, N)$ by eq.~(\ref{PsiScalingForm-recon}), and 
$S_{\rm av}(p, N)$ by eq.~(\ref{chiScalingForm}) 
of the configuration model.
$N=2^{20}$ (red-circle), $2^{19}$ (green-square), $2^{18}$ (blue-triangle), $2^{17}$ (open-diamond), and $2^{16}$ (open-circle).
}
\label{scaling-recon}
\end{figure*}


We show the scaling plot of $S_0 (p, N)$ and $\psi (p, N)$ in figs.~\ref{flowerdata2}(a) and (b), respectively, by using the exponents mentioned above. 
The collapsing of data to universal scaling function is very nice for data with large $N$ 
(as seen fig.~\ref{flowerdata1}(b), some correction to scaling cannot be neglected for $N < 10^{20}$).

\subsection{RAGN and configuration model}

The RAGN model stochastically generates a graph realization with $N$ nodes as follows. 
We start from a triangle. 
At each step, a new node joins and links to two randomly-chosen pre-existing nodes.
This process continues until the number of nodes reaches $N$.
The stationary degree distribution $P(k)$ of the resulting network is $P(k) \propto (2/3)^{k}$.

The rigorous result for the bond percolation on the RAGN is given by Bollob\'as and Riordan~\cite{riordan2005small,bollobas2005slow}: 
the transition point $p_{c2}$, above which a giant component having size of $O(N)$ emerges, is given by $p_{c2}=( 1-1/\sqrt{2} )/2$ 
and for $p>p_{c2}$ the order parameter follows eq.~(\ref{infiniteorder}) with $\alpha = \pi/2^{5/4}$ and $\beta'=1/2$. 
Also, we have $\psi_c = 1/2$ from the power-law behavior of $n_s$ at $p_{c2}$ (see eq.(6) in \cite{krapivsky2004universal}).

We also consider the configuration model having the same degree distribution as the RAGN. 
The bond percolation on the configuration model is analyzed by local tree approximation: 
the critical point, between the non-percolating phase and percolating phase, is given by $p_c=\langle k \rangle/\langle k^2-k \rangle=2/9$~\cite{cohen2000resilience}, 
and this transition belongs to the mean field universality class, $\beta=1$~\cite{cohen2002percolation}.

We performed Monte-Carlo simulations for bond percolation on the RAGN and the configuration model. 
For extensive simulations, we used Newman-Ziff algorithm \cite{newman2000efficient}.
The number of graph realizations is 1000, and the number of percolation trials on each realization is 100.
The order parameter $S_{\rm mac}(p, N)/N$ and the fractal exponent 
$\psi(p, N)\approx  \log_2 [S_{\rm max}(p, N) / S_{\rm max}(p, N/2)]$ on the RAGN are shown in fig.~\ref{rawdata}.
For each value of $p$ below $p_{c2}$, $\psi(p, N)$ almost converges to a certain value, 
while $\psi(p, N)$ for $p>p_{c2}$ approaches unity very slowly with $N$.
In spite of our extensive simulations, $\psi(p, N)$ at $p_{c2}$ is slightly smaller than $\psi_c$, 
due to the logarithmic correction in the power-law of $n_s$ \cite{krapivsky2004universal}.
But, $\psi(p, \infty)$ 
grows continuously with $p$ for $0(=p_{c1})<p<p_{c2}$ up to $\psi_c \simeq 1/2$ at $p_{c2}$, and then jump to $\psi=1$.
We obtain good data collapses for both $S_{\rm max}(p, N)$ and $\psi(p, N)$ by the finite size scaling, as shown in fig.~\ref{scaling}.
\footnote{
We did not perform the finite size scaling of $S_{\rm av}$ for the RAGN.
Because our numerical result for the RAGN shows $\psi_c=1/2$ at the transition point $p=p_{c2}$, leading to $\psi_{\rm av}=0$. 
This indicates that the mean cluster size of the RAGN does not diverge. 
Actually, a finite jump of $S_{\rm av}$ at $p_{c2}$ is observed on the RAGN~\cite{zalanyi2003properties}.
}.
Here parameters $\alpha$ and $\beta$ are set to the analytically obtained values.

The fractal exponent is useful not only to confirm the existence of the critical \textit{phase}, 
but also to determine the critical \textit{point} of ordinary continuous phase transition. 
In fig.~\ref{reconrawdata}, $S_{\rm max}(p, N)/N$, $\psi(p, N)$, and $S_{\rm av}(p, N)$ on the configuration model are plotted with respect to $p$.
In this model, $\psi(p, N)$ with several sizes cross at $(p_c, \psi_c) = (2/9, 2/3)$.
In the limit $N \to \infty$, $\psi=0(=1)$ for $p<p_c (>p_c)$, which means the transition at $p_c$ is the one between the non-percolating phase and the percolating phase.
The giant component just at the transition point is of $O(N^{2/3})$, which is also observed on the Erd\"os-R\'enyi model~\cite{janson1993birth}.
In fig.~\ref{scaling-recon}, we show the results for finite size scalings of $S_{\rm max}(p, N)$, $\psi(p, N)$, and $S_{\rm av}(p, N)$ with $\beta=1$.
Again, we can observe good data collapses for these measures, confirming the validity of our scaling hypothesis.

\section{Summary}

To summarize, we proposed a novel scaling analysis for critical phenomena in complex networks.
We confirmed the validity of our scaling via some exactly-solved models: 
the decorated (2,2)-flower, the RAGN and the configuration model.

Whether percolations on the other complex networks have such a critical phase or not is not known.
Also, whether other dynamics, e.g., epidemic spreading, coupled oscillators and so on, have such a critical phase or not is an interesting open problem.
By answering to these questions, the relation between network topology and the dynamics thereon will become clearer. 
We believe that the present method will contribute to further analyses.

\acknowledgments
The authors thank M. Sato, K. Konno, and N. Masuda for valuable comments.
TH acknowledges the support through Grant-in-Aid for Young Scientists(B) (No. 24740054) from MEXT, Japan. 
This research was partially supported by JST, ERATO, Kawarabayashi Large Graph Project. 


\end{document}